\documentclass[preprint]{vgtc}               

\ifpdf
  \pdfoutput=1\relax                   
  \usepackage{graphicx}                
  \DeclareGraphicsExtensions{.pdf,.png,.jpg,.jpeg} 
\else
  \usepackage{graphicx}                
  \DeclareGraphicsExtensions{.eps}     
\fi%

\graphicspath{{figures/}{pictures/}{images/}{./}}

\usepackage{microtype}                 
\PassOptionsToPackage{warn}{textcomp}  
\usepackage{textcomp}                  
\usepackage{mathptmx}                  
\usepackage{times}                     
         
\usepackage{cite}                      
\usepackage{tabu}                      
\usepackage{booktabs}                  
\usepackage{amsmath}

\usepackage{tabularx}
\usepackage{array}
\usepackage{multirow}
\usepackage{stfloats}
\usepackage{url}

\urldef{\urlML}\url{https://developer-docs.magicleap.cloud/docs/guides/features/eye-tracking/}

\onlineid{1038}

\vgtccategory{Research}

\vgtcinsertpkg

\preprinttext{This is the author’s version, to appear in the IEEE VR 2026 conference.}

\title{Evaluating the Viability of Additive Models to Predict Task Completion Time for 3D Interactions in Augmented Reality}

\author{Logan Lane\thanks{e-mail: logantl@vt.edu}\\ %
        \scriptsize Virginia Tech %
\and Ibrahim Tahmid\thanks{e-mail: iatahmid@vt.edu}\\ %
     \scriptsize Virginia Tech
\and Feiyu Lu\thanks{e-mail: feiyulu@vt.edu}\\ %
     \scriptsize Virginia Tech
\and Doug A. Bowman\thanks{e-mail: dbowman@vt.edu}\\ %
     \scriptsize Virginia Tech}

\teaser{
    \centering
  \includegraphics[width=0.78\linewidth]{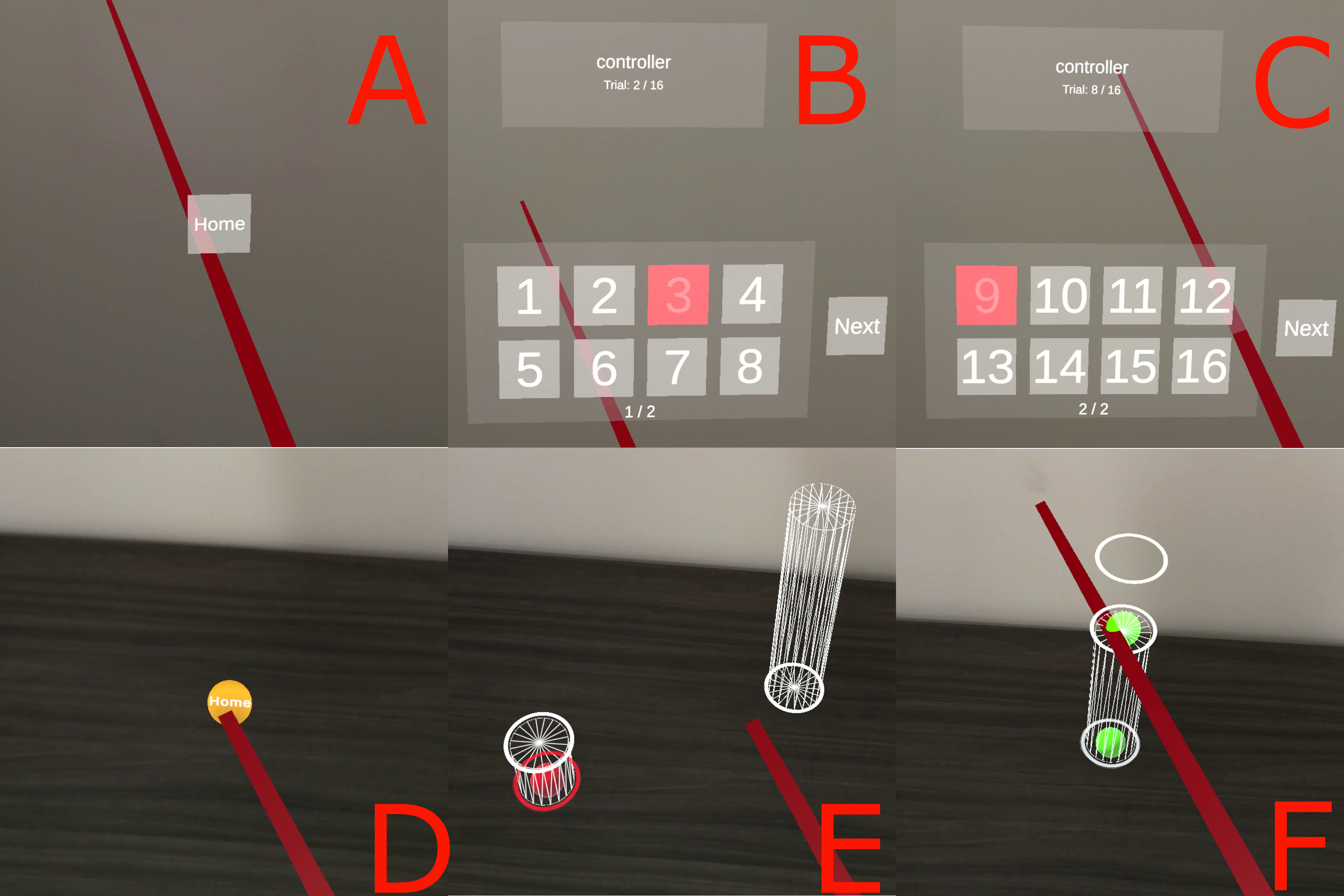}
  \vspace{-10pt}
  \caption{The interfaces used in the menu selection (A-C) and manipulation (D-F) studies. \textbf{A).} The home button that is pressed to display the menu. \textbf{B).} The first page of the menu. \textbf{C).} The second page of the menu. \textbf{D.)} The home sphere that is pressed to spawn the two cylinders shown in E. \textbf{E.)} The cylinder to be manipulated (left) and the target cylinder (right) at the beginning of the trial. \textbf{F).} The cylinder, after placement is complete and during the scaling phase of the trial.}
  \label{TeaserImage}
}

\abstract{Additive models of interaction performance, such as the Keystroke-Level Model (KLM), are tools that allow designers to compare and optimize the performance of user interfaces by summing the predicted times for the atomic components of a specific interaction to predict the total time it would take to complete that interaction. There has been extensive work in creating such additive models for 2D interfaces, but this approach has rarely been explored for 3D user interfaces. We propose a KLM-style additive model, based on existing atomic task models in the literature, to predict task completion time for 3D interaction tasks. We performed two studies to evaluate the feasibility of this approach across multiple input modalities, with one study using a simple menu selection task and the other a more complex manipulation task. We found that several of the models from the literature predicted actual task performance with less than 20\% error in both the menu selection and manipulation study. Overall, we found that additive models can predict both absolute and relative performance of input modalities with reasonable accuracy.
} 

\CCScatlist{
  \CCScatTwelve{Human-centered computing}{Human Computer interaction (HCI)}{Interaction paradigms}{Mixed / Augmented Reality};
  \CCScatTwelve{Human-centered computing}{Human Computer interaction (HCI)}{HCI design and evaluation methods}{User models};
}

\begin{document}


\firstsection{Introduction}

\maketitle
The Goals, Operators, Methods, and Selection Rules (GOMS) predictive model was proposed by Card et al. in 1983 \cite{cardPsychologyHumancomputerInteraction1983}. GOMS, and subsequent improved versions\cite{johnExtensionsGOMSAnalyses1990,grayProjectErnestineValidating, kierasGuideGOMSModel}, have been the standard for predicting how long an experienced user will take to complete a given interaction on a 2D user interface (UI). The earliest version of GOMS, called the Keystroke-Level Model (KLM) \cite{cardKeystrokelevelModelUser1980}, provided time predictions for simple 2D UIs by breaking an interaction task down to its atomic elements (or operators) and adding the predicted execution times for each to predict the total interaction time. An example of modeling with KLM is typing the word `cat'. Breaking this interaction down to its atomic operators, the user would first move a finger over the `c' key and then press it. This process would repeat for the other two letters, leaving us with three movement operators and then three keystrokes or confirmation operators. Summing the individual execution times for the six operators would provide a time prediction for the interaction. 

As an \textit{additive model}, KLM allows designers to quickly develop efficient interfaces by breaking them down to a sequence of elemental actions and estimating the time for each action. This approach saves time and resources, since the interfaces neither need to be fully implemented nor tested in a controlled user study to determine their efficiency. While most prior work on KLM has focused on 2D UIs, additive models for 3D UIs could similarly allow designers to predict task completion time (TCT) on multiple iterations of the same 3D UI. This could result in even more time savings due to the complexity of implementing and evaluating 3D UIs\cite{laviola20173d}. However, creating a 3D GOMS-style predictive model is not a straightforward endeavor. 3D UIs frequently involve fine-grained motor movements in 3D space with six Degrees of Freedom (DOF) controllers, hand, and eye gaze. This is further complicated by multi-modal interactions that use a combination of input modalities, such as gaze+pinch interactions\cite{pfeufferGazePinchInteraction2017}. Another barrier to a 3D GOMS-style additive model is the lack of standard predictive models for many atomic tasks with the aforementioned input modalities\cite{triantafyllidisChallengesModelingHuman2021, aminiSystematicReviewFitts2025}. In past research, extensive models have been proposed and tested for 3D movements such as controller pointing\cite{murataExtendingFittsLaw2001, kopperHumanMotorBehavior2010, teatherPointing3DTargets2011, laneRevisitingPerformanceModels2025}, bare-hand touching\cite{chaExtendedFittsLaw2013, zhaoMovementTimePointing2023}, and gaze pointing\cite{schuetzExplanationFittsLawlike2019, wagnerFittsLawStudy2023}; however, these models have had varying degrees of success in accurately predicting interaction time. 

The contribution of our work is two-fold. First, we explore the viability of using additive models to predict interaction completion time when performing menu selection and manipulation tasks using 3D user interfaces. We developed a KLM-style approach where an interaction is broken down to its atomic components, including movement time ($MT$) operators (the estimated time to move between a starting point and an end point) and confirmation operators (the time it takes to perform an action to signify the end of one phase of movement, such as a button press). Those individual time predictions are summed to provide an overall predicted $TCT$. While there have been similar attempts to use KLM-style models to predict the time to complete interactions on a 3D interface, our work features a much wider range of input modalities\cite{ghasemiEvaluatingUserInteractions2023, zhouHGOMSModelEvaluating2023}. A secondary contribution of our work is the validation of some predictive models of $MT$ operators and confirmation operators from past literature to see how accurate they are in the context of realistic 3D interaction tasks. We found that using additive models to predict $TCT$ for both simple and complex interactions is valid when used with some of the models from previous literature. Furthermore, we validated that the selected MT models yield accurate time predictions for Augmented Reality (AR) interactions. This is particularly significant as, to our knowledge, these models have not been previously assessed in an AR context, rather they were tested in either Virtual Reality (VR) or without immersive technology at all. Our work provides valuable insights into the application of KLM-style models for predicting the performance of 3D UIs when using a wide range of input modalities. 

\section{Related Work}
\subsection{Fitts' Law Predictive Models}
Fitts' Law was proposed by Paul Fitts during the 1950s and was able to successfully predict the amount of time it would take an individual to reciprocally move between physical plates of varying distance (A) and size (W)\cite{fittsInformationCapacityHuman1954}. Fitts' Law was defined as: 
\vspace{-5pt}
\begin{equation}
    MT_{Fitts} = a + b \cdot ID, \quad \text{where} \quad ID = \log_2\left(\frac{2A}{W}\right)
    \label{eq:Fitts}
\end{equation}

Fitts' Law played a key role in predictive systems like KLM being possible, as it allowed UX designers to estimate the time that it would take an expert user to complete a pointing-based interaction on a 2D user interface, serving as the pointing ($P$) operator in KLM\cite{cardKeystrokelevelModelUser1980, cardPsychologyHumancomputerInteraction1983}. 

There have been several attempts by researchers to extend Fitts' Law to work with different modalities, such as controller-based distal pointing\cite{murataExtendingFittsLaw2001, grossmanPointingTrivariateTargets2004, kopperHumanMotorBehavior2010, kopperRapidAccurate3D2011, laneRevisitingPerformanceModels2025} and gaze \cite{schuetzExplanationFittsLawlike2019, wagnerFittsLawStudy2023}. The approach has also been applied to direct touch interactions in 3D space \cite{chaExtendedFittsLaw2013, zhaoMovementTimePointing2023}. Applying Fitts' Law to 3D interactions is considered an open research question, since there is still no agreed-upon predictive model for 3D interactions as there is for 1D and 2D interactions \cite{teatherPointing3DTargets2011, triantafyllidisChallengesModelingHuman2021}. As a result, a contribution of our work is determining if the existing predictive models are sufficient for predicting menu selection and manipulation tasks in 3D space. 

The predictive models used in our work for the $MT$ operators add different parameters to the model based on the type of interaction they are trying to predict (e.g., target depth change for hand-based interactions\cite{barreramachucaEffectStereoDisplay2019}.) However, a commonality between all of the predictive models used for the $MT$ operators is that they use the original Fitts' law model as a starting point; that is, all of the models use the distance between a starting point and target, as well as the size of the target, to predict the amount of time an interaction will take.

\subsection{GOMS Models For 2D User Interfaces}
Card et al. developed the Goals, Operators, Methods, and Selection Rules (GOMS) predictive model\cite{cardPsychologyHumancomputerInteraction1983}. GOMS allows UX practitioners to estimate the amount of time that a specific interaction will take to complete by an experienced user. GOMS begins by looking at the specific goal to be completed with the software to be tested. Card et al. provide an example in their book in which GOMS is used to predict the amount of time to edit a manuscript\cite{cardPsychologyHumancomputerInteraction1983}. The overall goal, editing the manuscript, has multiple sub-goals, such as looking at the manuscript to determine that an edit needs to be made. Each sub-goal has one or more methods, comprised of multiple operators, that can be used to complete the sub-goal. Each operator is a specific action executed within the software and is one individual component of a method. Finally, it could be the case that there are multiple methods for accomplishing a goal. In this case, there are selection rules that can be used to determine which specific method should be used. Finally, the time from all operators is summed to provide a predicted time to complete an interaction. Past GOMS literature has used 20\% as the acceptable level of error between actual and predicted interaction time \cite{cardKeystrokelevelModelUser1980, johnCumulatingScienceHCI1989, olsonGrowthCognitiveModeling1995}.

Card et al. specifically proposed KLM in 1980\cite{cardKeystrokelevelModelUser1980}. KLM allowed designers to predict how long a given interaction would take by breaking the interaction down to its individual components. KLM is a simplified version of GOMS that does not use the more advanced operators offered by GOMS, instead looking at only keystrokes and other simple actions to predict interaction time \cite{johnWhyGOMS1995, cardPsychologyHumancomputerInteraction1983}. KLM used six operators to predict the completion time for an interaction: Keystroking ($K$), Pointing ($P$), Homing ($H$), Drawing ($D$), Mental ($M$), and System Response ($R$). Card et al. conducted a user study to test their predicted model and found that KLM had an error rate of around 20\%.

KLM has been a heavily researched topic for decades and has been expanded to include a wide variety of use cases \cite{al-megrenSystematicReviewModifications2018}, such as: mobile touch screen interfaces \cite{liExtendedKLMMobile2010, elbatranEnhancingKLMKeystrokelevel2014, holleisKeystrokelevelModelAdvanced2007, riceTouchlevelModelTLM2014, abdulinUsingKeystrokelevelModel2011}, in-vehicle infotainment systems\cite{pettittExtendedKeystrokeLevel2007}, and on-screen keyboards on televisions\cite{chatterjeeIterativeMethodolgyImprove2012}. Thus, we surmise that 3D interactions could successfully be modeled using a KLM-style approach as well. The approach followed in this paper is most similar to KLM in that we look at all the movements and confirmations (e.g., a button press or blink) required to complete a specific 3D interaction. The predicted times of the individual $MT$ and confirmation operators are then added together to predict the overall time to complete an interaction.

\subsection{Additive Models For 3D User Interfaces} 
There have been attempts to create extensions of GOMS that could closely predict interaction times in 3D space \cite{zhouHGOMSModelEvaluating2023, ghasemiEvaluatingUserInteractions2023}. Zhou et al. developed H-GOMS, a predictive model for 3D hand-based interactions\cite{zhouHGOMSModelEvaluating2023}. The model included operators for the actions performed by users as well as perception and cognition operators. The authors found their model to be accurate for expert users after conducting a set of studies to establish time baselines for the operators as well as another study to test the model itself. Ghasemi et al. used a GOMS-style model to predict TCT with gaze-pinch, gaze-voice, and pinch-drag interactions and concluded that the GOMS model was accurate enough to predict TCT \cite{ghasemiEvaluatingUserInteractions2023}. Our work differs from these studies in that we focus on a wide range of input modalities and seek to establish the validity of a variety of existing models. 

Other work is more closely aligned with KLM predictive modeling, where atomic individual operators are added together to get an overall prediction time for an interaction. Two previous works from the literature looked exclusively at creating new operators that represented common interactions in VR and AR, such as button presses and air taps \cite{guerraExtensionKeystrokelevelModel2022, cabricPredictivePerformanceModel2021}. Guerra et al. conducted a study that generated times for operators of common 3D interactions such as grabbing and teleporting \cite{guerraExtensionKeystrokelevelModel2022}. Similarly, Cabric et al. conducted a study that generated time values for new interaction operators\cite{cabricPredictivePerformanceModel2021}. They conducted another study to test the new time values in a basic KLM model where they found their operator time values to provide accurate predictions for the tested interactions. Overall, the literature suggests that performance for basic 3D interaction techniques such as pointing and selecting targets can be predicted successfully by using a KLM-style approach to modeling. However, little work has investigated complex 3D interaction tasks involving multiple modalities and precise manipulation of targets. While previous work has looked at modeling the time to complete gesture, voice, and pointing interactions for 3D UIs, our work seeks to further prove the feasibility of this idea for complex tasks, along with testing advanced input modalities that combine multiple types of input operators.  

\section{Model Components for Multiple Input Modalities}
The following subsections detail the predictive models from past literature that we used as $MT$ and confirmation operators in our additive models to make interaction time predictions. We intentionally only modeled the motor components of the interaction, avoiding the difficulty of modeling user perception and cognition in these initial studies of the approach. The selection of models and operators was guided by the following criteria: (1) reported accuracy, (2) recency of publication, and (3) frequency of adoption by other research. This work uses six input modalities: Distal pointing with discrete confirmation via trigger pull or blinking (\textsc{Controller}, \textsc{ControllerBlink}), gaze selection with trigger, airtap, or dwell confirmation (\textsc{GazeController}, \textsc{GazeAirtap}, and \textsc{GazeDwell}), and direct touch with pinch confirmation (\textsc{Hand}).
\subsection{Confirmation Operators}
The interaction operators proposed in a recent work by Cabric et al. were used as confirmation operators in our work\cite{cabricPredictivePerformanceModel2021}. These confirmation operators are denoted by $CO$. Specifically, the button press operator, with a value of 208 ms, was used for the \textsc{Controller} and \textsc{GazeController} input modalities. The airtap operator had a value of 428 ms and was used in the \textsc{GazeAirtap} modality. For the \textsc{Hand} modality, we used 214 ms for the pinch time since a pinch would have been half of the full 428 ms for the airtap operator. We also used the estimated time to complete a single eye blink of 200 ms from Caffier et al.'s work\cite{caffierExperimentalEvaluationEyeblink2003}. This was used in the \textsc{ControllerBlink} modality. We chose 500 ms as the dwell time to be used in the \textsc{GazeDwell} modality, as it has been widely used in prior work \cite{zhangModelingDwellbasedEye2010} to avoid the Midas Touch problem \cite{jacobWhatYouLook1990} while not requiring an overly lengthy dwell.

\subsection{Distal Pointing with Discrete Confirmation}
There have been many attempts to create a Fitts' Law-style predictive model to predict the movement time for distal pointing tasks on 3D interfaces. One of the earliest attempts at extending Fitts' Law was conducted by Murata and Iwase where they attempted to extend Fitts' law to support 3D distal selection tasks by adding a direction parameter to $ID$\cite{murataExtendingFittsLaw2001}. This provided a better fit of their data than the standard Fitts' law model, but was still a poor fit overall. Several other works attempted to modify the calculation of $ID$ to provide a better fit for 3D selection tasks by adding different parameters. Clark et al. added an inclination parameter to account for the incline of the user interface. Other researchers looked at how depth affected Fitts' law modeling performance as well as adding a depth parameter to $ID$\cite{kopperHumanMotorBehavior2010, teatherPointing3DTargets2011, teatherPointing3dTarget2013}. A recent work by Lane et al. has shown that using purely angular measurements in the $ID$ calculation provides an excellent fit when modeling 3D selection tasks with current VR hardware\cite{laneRevisitingPerformanceModels2025}. We chose to use this model due to its elegance and high level of fit to the data. The model used from Lane et al.'s work was defined as: 
\vspace{-10pt}
\begin{equation}
    MT_{DP} = a + b \cdot ID_{\text{ANG}}, \quad \text{where}  \quad ID_{\text{ANG}} = \log_2 \left( \frac{\alpha}{\omega} + 1 \right)
    \label{eq:IDANG}
\end{equation} 

$\alpha$ is the angular distance between the starting point and the destination target and $\omega$ is the angular size of the destination target. The constants $a$ and $b$ were empirically derived through regression. Using their study data, Lane et al. found that $a$ = 0.21 and $b$ = 0.16.  

This $ID_{ANG}$ model was used in two separate interaction modalities in our studies. Both interaction modalities had the participant use a controller with a ray emanating from the front. Participants pointed at the desired target and confirmed the selection by either pulling the trigger on the back of the controller (\textsc{Controller}) or blinking their eyes (\textsc{ControllerBlink}). It should be noted that we did not include a button press operator for the \textsc{Controller} modality, and we subtracted 208 ms from \textsc{ControllerBlink} movement time predictions. This was done because $ID_{ANG}$ already includes a trigger pull in its prediction. The equation for one phase of movement for the \textsc{Controller} modality is the same as Equation \ref{eq:IDANG}. Equation \ref{EQ:ControllerBlinkTrial} shows the structure for one phase of movement and confirmation for the \textsc{ControllerBlink} modality.
\vspace{-10pt}
\begin{equation}
\begin{aligned}
    Phase_{ControllerBlink} = MT_{DP} + CO_{Blink} - CO_{Trigger} \\ 
    \text{where}  \quad CO_{Blink} = 200ms \quad and \quad CO_{Trigger} = 208ms
\end{aligned}
\label{EQ:ControllerBlinkTrial}
\end{equation}

\subsection{Gaze}
There has been extensive prior work on gaze performance modeling with Fitts' Law type models, with work dating back to 1980s \cite{wareEvaluationEyeTracker1986}. However, some have argued that Fitts' Law alone is insufficient for modeling gaze performance due to the ballistic nature of gaze movements during saccades \cite{drewesOnlyOneFitts2010, schuetzExplanationFittsLawlike2019,drewesEyeGazeTracking}.

A recent work systematically evaluated gaze selection time\cite{schuetzExplanationFittsLawlike2019}. They tested the predictive capability of existing Fitts' Law models when gaze was used as a pointing device using the original model~\cite{fittsInformationCapacityHuman1954} and the Shannon formulation~\cite{mackenzieExtendingFittsLaw1992} by using a professional eye tracker and a 2D projected display to show the targets. They found that the original Fitts' Law was applicable when $ID$ was above a threshold, $ID_{crit}=1.74$. $ID_{crit}$ was determined by leaving out the smallest $ID$ values, followed by the next smallest $ID$ value, and so on until the largest $R^2$ value ($98.4\%$) was found. Practically, this means that movement time for any task with an $ID < 1.74$ can be determined by a constant value, the time it takes to complete one saccade of eye movement. This is due to the ballistic nature of initial eye movement saccades in which a given target can be gazed at using only one saccade from the starting point\cite{hoffmannCriticalIndexDifficulty2016, schuetzExplanationFittsLawlike2019}. Schuetz et al. found this constant value to be 232 ms in their work \cite{schuetzExplanationFittsLawlike2019}. 

Due to its recency, in-depth exploration, and empirical validity, we chose to use this model \cite{schuetzExplanationFittsLawlike2019}. For movements with $ID < 1.74$, we used 232 ms as $MT$. Otherwise, we used the original Fitts' law model (see Eq. \ref{eq:Fitts}). We applied this for all of our gaze-based interaction modalities. Equations \ref{EQ:GazeDwellTrial}, \ref{EQ:GazeAirtapTrial}, and \ref{EQ:GazeControllerTrial} show one phase of movement and confirmation for the \textsc{GazeDwell}, \textsc{GazeAirtap}, and \textsc{GazeController} modalities. 

\vspace{-17pt}
\begin{equation}
\begin{aligned}
    Phase_{GazeDwell}=MT + CO_{Dwell}, \quad
    \text{where} \quad CO_{Dwell}=500ms
\end{aligned}
\label{EQ:GazeDwellTrial}
\end{equation}

\vspace{-17pt}
\begin{equation}
\begin{aligned}
    Phase_{GazeAirtap} = MT + CO_{Airtap}, \quad
    \text{where}  \quad CO_{Airtap} = 428ms
\end{aligned}
\label{EQ:GazeAirtapTrial}
\end{equation}

\vspace{-17pt}
\begin{equation}
\begin{aligned}
    Phase_{GazeController} = MT + CO_{Trigger}, \quad 
    \text{where}  \quad CO_{Trigger} = 208ms
\end{aligned}
\label{EQ:GazeControllerTrial}
\end{equation}

\subsection{Direct Touch}
A recent survey by Amini et al. \cite{aminiSystematicReviewFitts2025} listed three models to be most widely applied by existing literature for modeling user interactions in 3D proximal spaces \cite{ murataExtendingFittsLaw2001, chaExtendedFittsLaw2013,barreramachucaEffectStereoDisplay2019}. Early work by Murata et al. explored the feasibility of Fitts' Law for modeling performance of touching targets arranged on a planar surface \cite{murataExtendingFittsLaw2001}. Machuca and Stuerzlinger studied the impact of depth on hand movement tasks along the Z axis \cite{barreramachucaEffectStereoDisplay2019}. Cha and Myung studied the impact of incorporating the inclination and azimuth angles of the targets into their model\cite{chaExtendedFittsLaw2013}. These models primarily involve spatial movements of hands that are restricted to certain surfaces or axes. The model we consider the most relevant is by Machuca and Stuerzlinger\cite{barreramachucaEffectStereoDisplay2019}, which modeled 3D movements of a wand to directly touch targets displayed by a 3D television. The model was defined as:
\vspace{-10pt}
\begin{equation}
    \text{MT}_{Hand} = a + b * ID + c * CTD
    \quad \text{where} \quad ID = \log_2(\frac{D}{W}+1)
    \label{eq:Hand_equation}
\end{equation}

$D$ represents the distance between the starting point and the target and $W$ represents the size of the target. $CTD$ represents the change in target depth and is measured in centimeters. The constants a, b, and c were all empirically derived through regression. Using their study data, they found that a = 167.6, b = 273.5, and c = 3.35. The authors found that their model provided an $R^2$ value of 98\%.

We used the original model as an $MT$ operator in the additive models used in both of our studies to predict the time it would take a participant to directly interact with a target using their right index finger. In the menu selection study, participants poked the desired target with their index finger (so no confirmation operator was used) while in the manipulation study, participants pinched the object that they wished to interact with, moved their hand to translate or scale the object, and then released the pinch to end the interaction. Eq. \ref{EQ:HandTrial} shows one phase of movement and confirmation for the \textsc{Hand} modality in the manipulation study.
\vspace{-7pt}
\begin{equation}
\begin{aligned}
    Phase_{Hand} = MT_{Hand} + CO_{Pinch/Release}, \\ 
    \text{where}  \quad CO_{Pinch/Release} = 214ms
\end{aligned}
\label{EQ:HandTrial}
\end{equation}

\section{Experiment}
\subsection{Goals}
We had two primary goals for the menu selection and manipulation studies. The first goal was to determine the feasibility of using additive models to predict the time to complete a complex interaction. The other goal was to test the accuracy and ability of past predictive models from the literature to predict movement times in selection and manipulation tasks when used as an $MT$ operator in additive models. We aimed to answer the following research questions:

\textbf{RQ1}: Is it feasible to use additive predictive models to predict the completion time of complex AR interactions using tasks similar to those found in real-world 3D user interfaces?

\textbf{RQ2}: To what extent are existing models for MT validated when used to predict performance for individual components of more complex AR interactions?

\subsection{Experiment Design}
In both studies, there was a single independent variable: the input modality used. The menu selection study tested six input modalities: \textsc{Controller}, \textsc{ControllerBlink}, \textsc{GazeController}, \textsc{GazeAirtap}, \textsc{GazeDwell}, and \textsc{Hand}. The manipulation study used the same input modalities except for \textsc{GazeDwell}. \textsc{GazeDwell} was not used in the manipulation study as manipulation is a continuous action, so it would be impossible to know when a user's gaze was dwelling on a target location. Input modalities were counterbalanced using a balanced Latin square. 

Both studies had task completion time ($TCT$) as the primary dependent variable. After completing all trials for each of the input modalities, the participants completed a NASA-TLX questionnaire gauging the perceived workload with that modality\cite{hartDevelopmentNASATLXTask1988}. After completing all modalities, participants ranked them from slowest to fastest based on perceived speed.

\subsection{Apparatus}
For both studies, we used a Magic Leap 2 (ML2) AR Head-Worn Display (HWD) and controller that allowed participants to see and interact with AR elements in the environment. The study software for ML2 was built with the Unity 3D game engine (Version 2022.3.17f1) as well as the ML2 SDK from Magic Leap. 

In the manipulation study only, we had participants use the overhead strap in an attempt to resolve some eye-tracking issues we encountered in the menu selection study that could have been caused by the HWD slipping over the course of the study. We also wanted to ensure the HWD stayed in place, since participants would be looking down for the entirety of the manipulation study, which could make the HWD more likely to shift on the participant's head. The manipulation study also used a motorized standing desk. 

\subsection{Procedure}
Both studies were approved by our Institutional Review Board. Participants were emailed a consent form after they expressed interest in participating in the study. Upon arriving, participants signed the consent form acknowledging their willingness to participate. Next, participants were shown a presentation introducing them to AR and the input modalities that would be tested during the study. 

After the presentation, participants were shown how to wear the headset as well as the buttons on the controller that would be used throughout the study. Participants then donned the ML2 HWD and began the calibration process for headset fit and eye tracking. Upon completing the calibration, the experimenter guided participants to open the study application.

\subsubsection{Menu Selection Study}
During the menu selection study, participants selected buttons on a virtual menu. The menu (shown in Figure \ref{TeaserImage}) had two pages, with each page having eight buttons arranged in a 2x4 layout for a total of 16 buttons. Prior to beginning the trials, participants were instructed to calibrate the menu position by stretching their arm out directly in front of them so that the controller pointed away from their torso and pressing the bumper button. This made the menu appear slightly ahead of the controller, with the menu rotated to face the participant. 

Before beginning the trials for each input modality, participants completed a practice set of trials in order to become familiar with how the input modality worked. To complete a menu selection trial, participants used the current input modality to select a virtual home button that made the menu appear. The participant then selected the button highlighted in red. Buttons were highlighted in order from 1-16 in subsequent trials. This was done so that participants would not need to spend time searching for the next button to be selected, which would affect motor performance.  Buttons 9-16 required participants to select the ``Next'' button that was placed to the right of the menu. This showed the second page of the menu with the highlighted button. If the highlighted button was on the first page, only one phase of movement and confirmation was needed to complete a trial, while the second page required two phases, one to select the Next button and the other to select the target button. Note that the predicted time for selecting each menu item was computed independently and compared to the actual time for that selection. The menu was 14.16 inches wide and 7.44 inches tall. Each button was 2.28 inches tall and wide. The distance between two consecutive buttons was 3.19 inches. After calibration of the menu position, the average distance of the menu from the participant ranged from 1.8 to 3.18 feet with an average distance of 2.73 feet.

Participants were instructed to complete the formal trials as quickly as possible without making mistakes. They completed five sets of menu selection trials for a total of 80 trials. After the completion of each formal trial, the time the participant took to complete the trial was logged, as well as a predicted time that was computed for each individual trial using the additive models. After completing all trials, participants completed a NASA-TLX survey for the input modality that they had just used\cite{hartDevelopmentNASATLXTask1988}. Participants then moved on to the next input modality and completed another five sets of menu selection trials. This process was repeated until the participants completed five sets of trials for all six interaction modalities. After all the trials for each input modality were completed, the participant ranked each input modality based on perceived speed. Finally, participants completed a background questionnaire that asked them about age, occupation, dominant hand, and past VR/AR usage. 

\subsubsection{Manipulation Study}
Participants followed a similar procedure for the manipulation study. Prior to calibration, the table height was adjusted to match the height of the participant's waist. To calibrate, the controller was placed on the table at a marked location and the controller bumper was pressed to align an invisible, virtual table with the physical one, so that the virtual objects appeared on top of the physical table.

In a manipulation trial, participants first selected the home sphere (see \autoref{TeaserImage} D), which spawned a manipulation cylinder and a target cylinder. To systematically generate the cylinders, we designated four positions, two on each side of the home sphere, as potential starting locations for the manipulation cylinder. 

Because of the tilted viewing angle of the tabletop from the participant's perspective, the angular distance between the two cylinders depends on their spatial positions on the table, even when their linear distances do not vary, introducing potential bias during translations.
Therefore, for each starting location, we systematically placed $8$ target cylinders towards the far and near edges of the table that prompted depth-oriented translations and $8$ target cylinders towards the east or west edge that prompted side-to-side translations.
The linear distance between the manipulation cylinder and the $16$ target cylinders ranged from $0.12$ m to $0.35$ m, and the scale factor varied from $1.2$ to $4$.
Both distance and scale factors were randomized and seeded by input modality to ensure that all participants within a modality experienced the same sequence of (manipulation, target) pairs.
We repeated the same process for all 4 starting locations, leaving us with $64$ target cylinders.

Once the cylinders (see Figure \ref{TeaserImage}) appeared, the participant used the current input modality to select the manipulation cylinder and translate it along the table surface until it was inside the target cylinder. The participant then selected the handle on top of the manipulation cylinder and pulled it upwards until both cylinders had the same height. Each trial required three phases of movement and confirmation to complete: translating the manipulation cylinder (\textit{Placement Time}), moving to and selecting the handle (\textit{Scale Search Time}), and translating the handle (\textit{Scale Time}). Participants completed 64 trials for each input modality. The trial time began after the participant began the placement phase of the trial. In both studies, we computed the angular distances, $\alpha$ and $\omega$, by using either the position of the controller, hand, or eye at the time of pressing the Home button as the origin point.

\subsection{Participants}
We recruited 33 participants for the menu selection study and 22 participants for the manipulation study. Each study had its own participant pool with no overlap between the two.  Due to eye tracking instability, HWD fit issues, and procedure violations, we removed 9 participants from the menu selection study and 2 participants from the manipulation study. This left us with 24 participants (4 Female) in the menu selection study and 20 (8 Female) in the manipulation study, with ages ranging from 19 to 30 years old (mean=22). All but three participants had used immersive technology at least once with most having used immersive technology more than 10 times. All participants had normal or corrected (contact lenses only) vision.

\section{Results}
\subsection{Menu Selection Study Results}
\subsubsection{Additive Model Accuracy}
Each participant completed 480 trials with six input modalities. In total, there were 11,520 menu selection trials completed across the entire study. Across all modalities, participants failed 337 trials. Failed trials occurred when the participant selected the wrong button and were excluded from analysis, then immediately attempted again. For each modality individually, error rates ranged from 0.36\% (\textsc{Controller}) to 10.47\% (\textsc{Hand}). 

Following the example of prior work, any trial that had a delta time (actual time - predicted time) more than two standard deviations above or below the mean delta time were removed prior to beginning data analysis\cite{grossmanProbabilisticApproachModeling2005, kopperHumanMotorBehavior2010, laneRevisitingPerformanceModels2025}. We felt justified in removing outliers with this strategy for practical reasons. For example, we noticed that participants would sometimes struggle to use the hand tracking on the ML2, leading to longer trial times that were not truly representative of pure motor performance. Following this strategy, we removed 522 outliers, leaving us with 10,998 trials for analysis.

To understand the predictive capability of the additive models, we relied on three primary data analysis strategies. First, we looked at the percentage differences between the total predicted and actual times for each condition. This helped us to understand at a high level how different our predicted and actual datasets were. Next, we used a paired Z-Test to determine whether the actual and predicted value datasets were statistically different from one another. Each trial had its own predicted time value that was computed and logged, and an actual time value that was logged when the trial was completed. The Z-score represents the level of difference between the two groups. A p-value below the significance level of 0.05 indicates that there is a statistically significant difference between the two groups. We also computed the Cohen's D value for each significant difference to measure the effect sizes, which would provide evidence for how practically important the differences were.

Finally, we performed a Two One-Sided t-test (TOST) to examine equivalence between the actual and predicted times. To do this, we first computed the percentage difference between the actual and predicted times for each trial. We calculated the average percentage difference along with the standard deviation and defined upper and lower equivalence bounds (+20\% and -20\%, respectively\cite{cardKeystrokelevelModelUser1980, johnCumulatingScienceHCI1989, olsonGrowthCognitiveModeling1995}) that would be used to determine the level of acceptable difference to be considered practically equivalent. Finally, we conducted the TOST which returned a p-value. If the p-value returned was below the 0.05 significance level, then the actual and predicted times for a specific input modality could be considered equivalent within 20\%. Conducting all three analyses provided us with the evidence needed to determine whether a model was performing well.

\begin{figure}[ht]
    \centering
    \includegraphics[width=0.75\columnwidth]{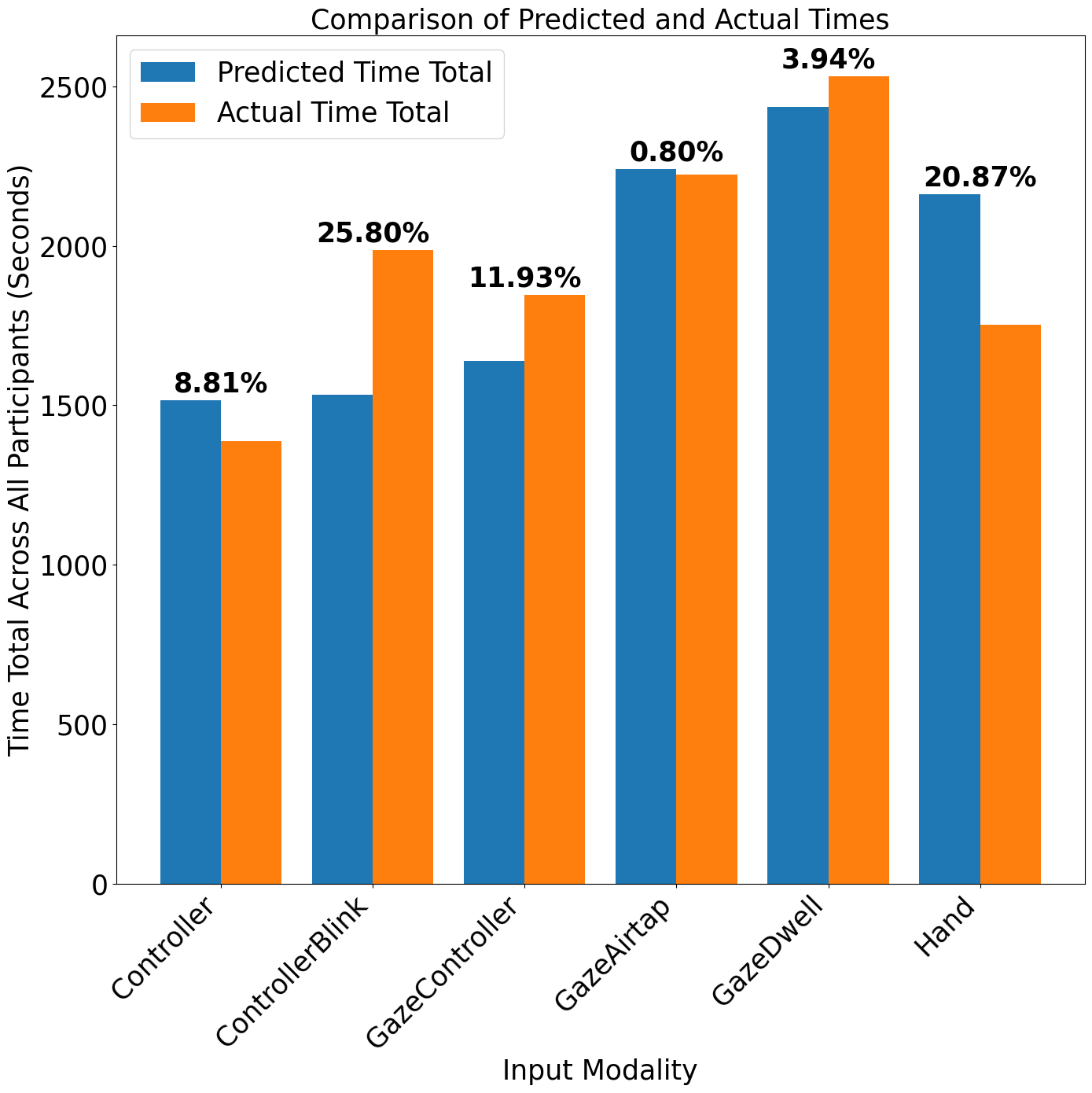}
    \vspace{-10pt}
    \caption{Total predicted and actual time with confirmation operators in seconds for all participants for all input modalities for the menu selection study.}
    \label{Totalvpredicted}
    \vspace{-20pt}
\end{figure}
Figure \ref{Totalvpredicted} shows the total actual and predicted time values across all of the menu selection trials and the percentage difference between the two. Figure \ref{Totalvpredicted} shows that two input modalities, \textsc{GazeAirtap} and \textsc{Controller}, beat the average time provided by their respective models. The prediction for \textsc{GazeAirtap} was nearly perfect with a percentage difference of 0.8\%, while the predictions for \textsc{ControllerBlink} were the least accurate, differing by 25.80\%.

Table \ref{MenuAverageValues} shows the average actual and predicted time values, along with the actual, predicted, and participant-perceived speed rankings.
The average difference between the predicted and actual rankings was 1. \textsc{GazeAirtap} and \textsc{GazeDwell} were correctly predicted to be the two slowest modalities. The other four modalities were close in their actual and predicted times, with all modalities being one rank off except for the \textsc{ControllerBlink} and \textsc{Hand} modalities, which were three and two ranks off, respectively. The perceived rankings matched the actual rankings for \textsc{Controller} (fastest) and \textsc{GazeAirtap} (second slowest). Outside of this, however, the perceived rankings were fairly different from the actual rankings. Another way to look at prediction performance is to consider pairwise predictions between modalities and whether the models correctly predict which modality in a pair will have better performance than the other. With six modalities, there are 15 unique pairs for comparison. Our additive models correctly predicted the better performing modality in 11 out of 15 pairs (73.3\%).

\begin{table}[ht]
\centering
\small 
\renewcommand{\arraystretch}{0.95}
\setlength{\tabcolsep}{5pt}
\caption{Predicted vs. actual vs. perceived modality rankings in the menu selection study.}
\begin{tabular}{l ccccc}
\toprule
Modality & \begin{tabular}[c]{@{}c@{}}Predicted \\ Average\\ Time\end{tabular} 
         & \begin{tabular}[c]{@{}c@{}}Predicted \\ Rank\end{tabular} 
         & \begin{tabular}[c]{@{}c@{}}Actual \\ Average\\ Time\end{tabular} 
         & \begin{tabular}[c]{@{}c@{}}Actual \\ Rank\end{tabular} 
         & \begin{tabular}[c]{@{}c@{}}Perceived\\ Rank\end{tabular} \\
\midrule
ControllerBlink  & 0.83 & 1 & 1.07 & 4 & 3 \\
Controller       & 0.84 & 2 & 0.76 & 1 & 1 \\
GazeController   & 0.89 & 3 & 1.00 & 3 & 2 \\
Hand             & 1.19 & 4 & 0.96 & 2 & 6 \\
GazeAirtap       & 1.22 & 5 & 1.21 & 5 & 5 \\
GazeDwell        & 1.33 & 6 & 1.38 & 6 & 4 \\
\bottomrule
\end{tabular}
\label{MenuAverageValues}
\vspace{-10pt}
\end{table}

The results of the paired Z-test, Cohen's D, and TOST are shown in Table \ref{tab:pairedztest-tostmenuwithconfirmationoperators}. Except \textsc{GazeAirtap} (p = 0.38), all modalities showed a statistically significant difference (p $<$ 0.001). We found that all modalities had a small effect size with the exception of \textsc{Hand} and \textsc{ControllerBlink}, which both had medium effect sizes (D = -0.63 and D = 0.78, respectively). The TOST showed the predictions and actual times for all input modalities except \textsc{ControllerBlink} can be considered statistically equivalent within 20\%.

\begin{table*}[!t]
  \centering
  \caption{Paired Z-Test and TOST results for the menu selection study. This data includes confirmation operators. Effect sizes (Cohen’s d) are indicated by daggers: no dagger for $|d|<0.2$ (very small/negligible effect); \dag\ for $0.2\le|d|<0.5$ (small); \dag\dag\ for $0.5\le|d|<0.8$ (medium); and \dag\dag\dag\ for $|d|\ge0.8$ (large). CI values represent the 90\% confidence interval bounds from the TOST analysis.}
  \small 
  \setlength{\tabcolsep}{4pt} 
  \renewcommand{\arraystretch}{0.95}
  \begin{tabular*}{\textwidth}{@{\extracolsep{\fill}} l cccc ccccc}
    \toprule
    & \multicolumn{4}{c}{Paired Z-Test} & \multicolumn{5}{c}{TOST Results} \\
    \cmidrule(lr){2-5} \cmidrule(lr){6-10}
    Modalities
      & $Z$ & $p$ & $SD$ & $d$
      & $t$ & $p$ & $SD$ & CI$_{\text{low}}$ & CI$_{\text{high}}$ \\
    \midrule
    Controller        & $-16.49$ & $\mathbf{<.001}$ & 0.18 & $-0.39$\dag
                      & 23.87    & $\mathbf{<.001}$ & 0.22 & $-0.09$ & $-0.07$ \\
    ControllerBlink   & 33.65    & $\mathbf{<.001}$ & 0.31 &  0.78\dag\dag
                      & 12.39    & 1                & 0.42 & 0.30  & 0.34 \\
    GazeAirtap        & $-0.87$  & 0.38             & 0.47 & $-0.02$
                      & $-19.50$ & $\mathbf{<.001}$ & 0.43 & $-0.01$ & 0.02 \\
    GazeController    & 12.89    & $\mathbf{<.001}$ & 0.38 &  0.30\dag
                      & $-5.17$  & $\mathbf{<.001}$ & 0.46 & 0.13  & 0.16 \\
    GazeDwell         &  5.73    & $\mathbf{<.001}$ & 0.40 &  0.13
                      & $-21.50$ & $\mathbf{<.001}$ & 0.32 & 0.03  & 0.05 \\
    Hand              & -27.04    & $\mathbf{<.001}$ & 0.35 &  -0.63\dag\dag
                      & 7.65    & $\mathbf{<.001}$                & 0.30 & -0.16  & -0.13 \\
    \bottomrule
  \end{tabular*}
  \label{tab:pairedztest-tostmenuwithconfirmationoperators}
  \vspace{-5pt}
\end{table*}

\subsubsection{Individual Model Accuracy}
We also analyzed the performance of the individual models to understand how the performance of the individual $MT$ and confirmation operators affected the performance of the additive model. We generated log files recording user actions throughout the menu selection study. To get the actual time for the individual $MT$ operators, we looked at the time where the participant entered the target button prior to performing the confirmation operator. The actual time for the confirmation operator was found by subtracting the logged time for a ``Trial End" event from the time logged when the participant last entered the target button. The predicted times for $MT$ were computed using their respective models. The confirmation operator constants from Cabric et al. and Caffier et al.'s works were used as the predicted times for the confirmation operators\cite{cabricPredictivePerformanceModel2021, caffierExperimentalEvaluationEyeblink2003}. 

The average time it took for participants to complete the confirmation operators was also computed. In comparing the actual and predicted time ($\Delta$ = Actual confirmation time - predicted confirmation time), we found that it took participants 119 ms ($\Delta$ = -89 ms) to complete a trigger pull in the \textsc{Controller} modality, 203 ms ($\Delta$ = 3 ms) to complete a blink in the \textsc{ControllerBlink} modality, 328 ms ($\Delta$ = 120 ms) to complete a trigger pull in the \textsc{GazeController} modality, and 418 ms ($\Delta$ = -10 ms) to complete an airtap in the \textsc{GazeAirtap} modality. The \textsc{Hand} modality does not have a formal confirmation mechanism as the selection occurs when the index finger's collider touches the button.

\begin{table*}[!t]
  \centering
    \caption{Paired Z-Test and TOST results for the menu selection study. This data does not include confirmation operators.}
  \small
  \setlength{\tabcolsep}{4pt}
  \renewcommand{\arraystretch}{0.95}
  \begin{tabular*}{\textwidth}{@{\extracolsep{\fill}} l cccc ccccc}
    \toprule
    & \multicolumn{4}{c}{Paired Z-Test Results} & \multicolumn{5}{c}{TOST Results} \\
    \cmidrule(lr){2-5} \cmidrule(lr){6-10}
    Modalities
      & $Z$ & $p$ & $SD$ & $d$
      & $t$ & $p$ & $SD$ & CI$_{\text{low}}$ & CI$_{\text{high}}$ \\
    \midrule
    Controller        & 17.36    & $\mathbf{<.001}$ & 0.12 & 0.33\dag
                      & $-11.38$ & $\mathbf{<.001}$ & 0.34 & 0.11 & 0.14 \\
    ControllerBlink   & 41.99    & $\mathbf{<.001}$ & 0.19 & 0.80\dag\dag\dag
                      & 23.46    & 1                & 0.58 & 0.44 & 0.48 \\
    GazeAirtap        & $-11.64$ & $\mathbf{<.001}$ & 0.23 & $-0.23$\dag
                      & 5.16     & $\mathbf{<.001}$ & 0.57 & $-0.16$ & $-0.12$ \\
    GazeController    & $-16.21$ & $\mathbf{<.001}$ & 0.21 & $-0.32$\dag
                      & 2.58     & $\mathbf{<.001}$ & 0.50 & $-0.19$ & $-0.16$ \\
    GazeDwell         & 6.93     & $\mathbf{<.001}$ & 0.34 & 0.14
                      & $-5.69$  & $\mathbf{<.001}$ & 0.83 & 0.08 & 0.13 \\
    \bottomrule
  \end{tabular*}
  \label{tab:pairedztest-tostmenuwithnoconfirmationoperators}
  \vspace{-12pt}
\end{table*}

To evaluate the performance of the \textit{MT} predictions alone, we removed the time it took to perform the confirmation operators from the predicted and actual interaction times per trial, leaving only the $MT$ value.) A paired Z-test, shown in Table \ref{tab:pairedztest-tostmenuwithnoconfirmationoperators}, found similar results as those in Table \ref{tab:pairedztest-tostmenuwithconfirmationoperators}. All modalities' predicted and actual times were statistically different (p $<$ 0.001) and had small effect sizes, except \textsc{ControllerBlink} which had a large effect size (D = 0.80). We also conducted a TOST on the MT operators and found that all modalities showed equivalence (within 20\%) between their actual and predicted time values except for \textsc{ControllerBlink}.

\subsection{Manipulation Study Results}
Each participant completed a total of 320 trials with five input modalities. In total, there were 7680 manipulation trials completed across the entire study. Across all modalities, participants failed 1055 trials with an overall error rate of 14\%. Failed trials were immediately attempted again and were not included for analysis. For each modality individually, the error rates ranged from 3.96\% (\textsc{Controller}) to 24.90\% (\textsc{GazeAirtap}). The high error rate for the \textsc{GazeAirtap} modality likely stems from participants struggling with the ML2 HWD's hand tracking as participants would sometimes have to ``airtap'' multiple times which would often lead to participants looking away from the target when the airtap finally registered, failing the trial.
Any trial that had a delta time (actual time - predicted time) more than two standard deviations above or below the mean delta time were removed for each of the modalities. Doing this removed 1521 trials and left 6159 trials for analysis. 
\vspace{-5pt}
\subsubsection{Additive Model Accuracy}
\vspace{-5pt}
\begin{figure}[ht]
    \centering
    \includegraphics[width=0.75\columnwidth]{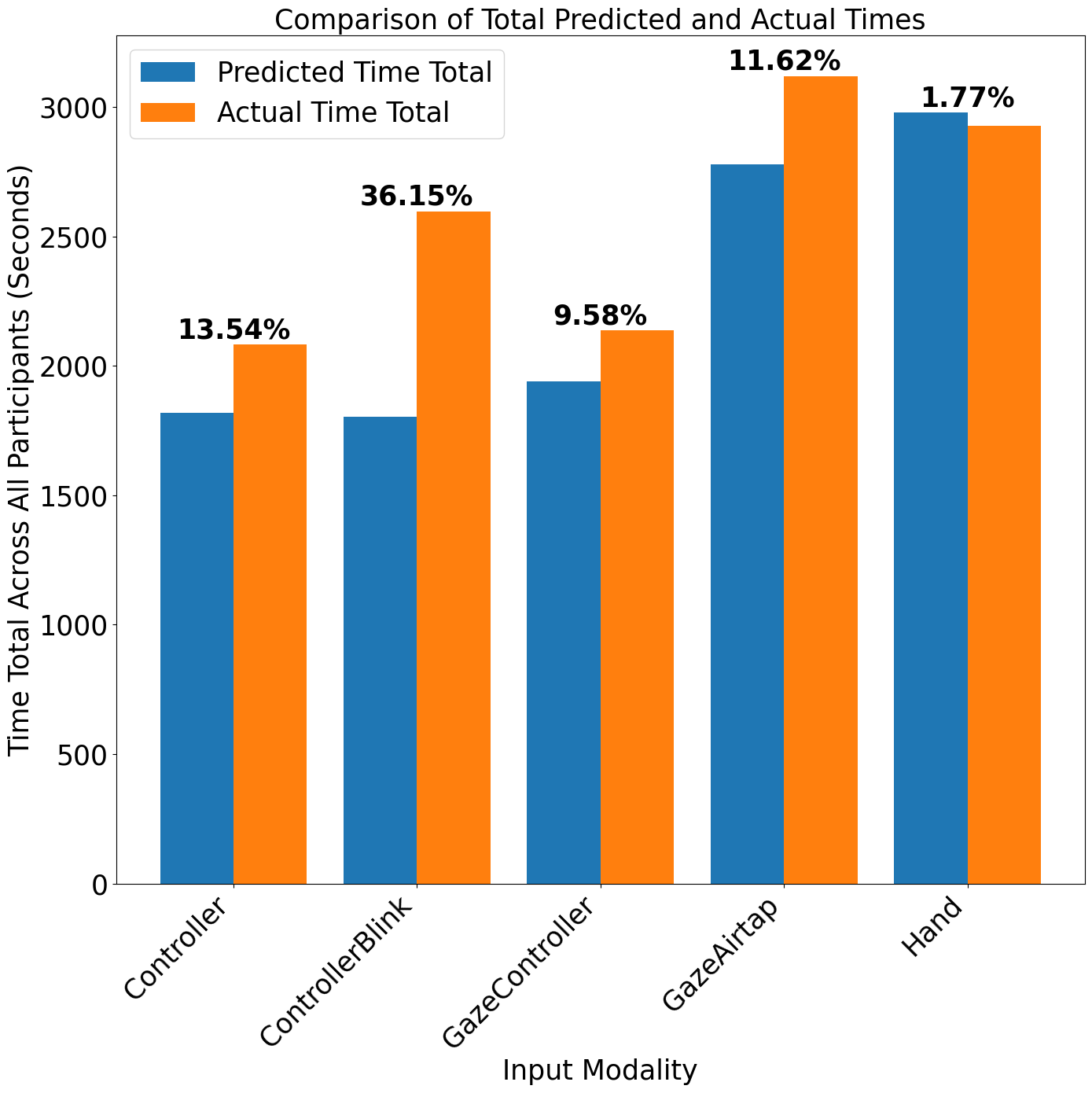}
    \vspace{-10pt}
    \caption{Total predicted and actual time for all participants for all input modalities in the manipulation study.}
    \label{TotalvpredictedManipulation}
    \vspace{-8pt}
\end{figure}

\begin{table}[ht]
\centering
\caption{Predicted vs. actual vs. perceived modality rankings in the manipulation study.}
\small 
\renewcommand{\arraystretch}{0.95} 
\setlength{\tabcolsep}{5pt} 
\begin{tabular}{l ccccc}
\toprule
Modality & \begin{tabular}[c]{@{}c@{}}Predicted \\ Average \\ Time\end{tabular} 
         & \begin{tabular}[c]{@{}c@{}}Predicted \\ Rank\end{tabular} 
         & \begin{tabular}[c]{@{}c@{}}Actual \\ Average \\ Time\end{tabular} 
         & \begin{tabular}[c]{@{}c@{}}Actual \\ Rank\end{tabular} 
         & \begin{tabular}[c]{@{}c@{}}Perceived \\ Rank\end{tabular} \\
\midrule
ControllerBlink & 1.46 & 1 & 2.10 & 3 & 3 \\
Controller      & 1.48 & 2 & 1.69 & 1 & 2 \\
GazeController  & 1.59 & 3 & 1.75 & 2 & 1 \\
GazeAirtap      & 2.24 & 4 & 2.52 & 5 & 5 \\
Hand            & 2.41 & 5 & 2.37 & 4 & 4 \\
\bottomrule
\end{tabular}
\label{ManipulationAverageValues}
\vspace{-22 pt}
\end{table}

Figure \ref{TotalvpredictedManipulation} shows the total actual and predicted time values across all the manipulation trials along with the percentage difference between the two time values for each input modality. Table \ref{ManipulationAverageValues} shows the average actual and predicted time values, along with the actual, predicted, and participant-perceived speed rankings. Both Figure \ref{TotalvpredictedManipulation} and Table \ref{ManipulationAverageValues} show that the actual time values were greater than the predicted for all but one of the input modalities (\textsc{Hand}). Taking the difference between the actual and predicted rankings for each modality gives an average difference of 1.2 ranks. For each of the individual modalities, we were able to predict the rankings within one rank of their respective actual rankings, except the \textsc{ControllerBlink} modality, which was two ranks away. Interestingly, the perceived ranks aligned very closely with the actual ranks, with \textsc{Controller} and \textsc{GazeController} being the only two differences. In terms of pairwise predictions, our additive models predicted the better performing modality in 7 of 10 pairs (70\%).

A paired Z-test was conducted between the predicted and actual times (shown in Table \ref{tab:pairedztest-tostmanipulationwithconfirmationoperators}). All five modalities showed a strong statistically significant difference (p $<$ 0.001). We found that \textsc{GazeController}, \textsc{GazeAirtap}, and \textsc{Hand} had small effect sizes (D = 0.30, D = 0.34, and D = -0.06, respectively), \textsc{Controller} had  a medium effect size (D = 0.50), and \textsc{ControllerBlink} had a large effect size (D = 1.26). We conducted a TOST between the predicted and actual times and found that the actual and predicted times for all input modalities except \textsc{ControllerBlink} were statistically equivalent within 20\%. 

\begin{table*}[!t]
  \centering
    \caption{Paired Z-Test and TOST results for the manipulation study. This data includes confirmation operators.}
  \small
  \setlength{\tabcolsep}{4pt}
  \renewcommand{\arraystretch}{0.95}
  \begin{tabular*}{\textwidth}{@{\extracolsep{\fill}} l cccc ccccc}
    \toprule
    & \multicolumn{4}{c}{Paired Z-Test Results} & \multicolumn{5}{c}{TOST Results} \\
    \cmidrule(lr){2-5} \cmidrule(lr){6-10}
    Modalities
      & $Z$ & $p$ & $SD$ & $d$
      & $t$ & $p$ & $SD$ & CI$_{\text{low}}$ & CI$_{\text{high}}$ \\
    \midrule
    Controller        & 17.44    & $\mathbf{<.001}$ & 0.43 & 0.50\dag\dag
                      & $-5.92$  & $\mathbf{<.001}$ & 0.30 & 0.14 & 0.16 \\
    ControllerBlink   & 44.40    & $\mathbf{<.001}$ & 0.51 & 1.26\dag\dag\dag
                      & 24.04    & 1                & 0.38 & 0.44 & 0.48 \\
    GazeAirtap        & 11.81    & $\mathbf{<.001}$ & 0.82 & 0.34\dag
                      & $-7.27$  & $\mathbf{<.001}$ & 0.37 & 0.11 & 0.14 \\
    GazeController    & 10.65    & $\mathbf{<.001}$ & 0.52 & 0.30\dag
                      & $-10.24$ & $\mathbf{<.001}$ & 0.33 & 0.09 & 0.12 \\
    Hand              & -2.15    & 0.03 & 0.69 & -0.06
                      & 22.59    & $\mathbf{<.001}$                & 0.29 & -0.03 & -0.0005 \\
    \bottomrule
  \end{tabular*}
  \label{tab:pairedztest-tostmanipulationwithconfirmationoperators}
  \vspace{-5pt}
\end{table*}

\subsubsection{Individual Model Accuracy}
Like in the menu selection study, we were interested in the individual performances of the $MT$ and confirmation operator models to better understand how they affected the performance of the additive model.
Due to a data logging error, we did not have separate time values for the movement and confirmation times for the placement and scaling phases. Instead, we had one time value that included both movement and confirmation time together. We did have separate movement and confirmation times for the search phase. The average time to complete the confirmation operator during the search phase across all participants was subtracted from the combined time values to get an approximate $MT$ for the placement and scaling phases.

The average time to perform the confirmation operators was computed for each modality ($\Delta$ = Actual confirmation time - predicted confirmation time). We found that it took participants 136 ms ($\Delta$ = -72 ms) to complete a trigger pull in the \textsc{Controller} modality, 268 ms ($\Delta$ = 68 ms) to complete a blink in the \textsc{ControllerBlink} modality, 329 ms ($\Delta$ = 121 ms) to complete a trigger pull in the \textsc{GazeController} modality, 541 ms ($\Delta$ = 113 ms) to complete an airtap in the \textsc{GazeAirtap} modality, and 345 ms ($\Delta$ = 131 ms) to complete a pinch/release in the \textsc{Hand} modality.

We conducted a paired Z test and a TOST on the three combined $MT$ operators without confirmation operators to gauge the individual performance of the $MT$ operators in the manipulation study. In this analysis, the average actual confirmation time from the search phase was removed from all three phases to get our approximate $MT$ model times for the three phases. The results of the test are reported in Table \ref{tab:pairedztest-tostmanipulationwithnoconfirmationoperators}. We found that four of the five modalities showed a strong statistically significant difference between the predicted and actual time values (p $<$ 0.001) with \textsc{GazeAirtap} showing no significant difference (p = 0.29). \textsc{GazeController} and \textsc{GazeAirtap} had negligible or small effect sizes, \textsc{Hand} had a medium effect size while \textsc{Controller} and \textsc{ControllerBlink} had a large effect size. The results from the TOST showed that only the \textsc{GazeAirtap} modality showed significant equivalence within 20\%.

\begin{table*}[!t]
  \centering
    \caption{Paired Z-Test and TOST results for the manipulation study using only the combined $MT$ operators for each modality. This data does not include confirmation operators.}
  \small
  \setlength{\tabcolsep}{4pt}
  \renewcommand{\arraystretch}{0.95}
  \begin{tabular*}{\textwidth}{@{\extracolsep{\fill}} l cccc ccccc}
    \toprule
    & \multicolumn{4}{c}{Paired Z-Test Results} & \multicolumn{5}{c}{TOST Results} \\
    \cmidrule(lr){2-5} \cmidrule(lr){6-10}
    Modalities
      & $Z$ & $p$ & $SD$ & $d$
      & $t$ & $p$ & $SD$ & CI$_{\text{low}}$ & CI$_{\text{high}}$ \\
    \midrule
    Controller        & 39.78    & $\mathbf{<.001}$ & 0.40 & 1.13\dag\dag\dag
                      & 24.13    & 1                & 0.53 & 0.54 & 0.59 \\
    ControllerBlink   & 35.26    & $\mathbf{<.001}$ & 0.46 & 1.00\dag\dag\dag
                      & 21.83    & 1                & 0.67 & 0.59 & 0.65 \\
    GazeAirtap        & 1.05     & 0.29             & 0.75 & 0.03
                      & $-7.87$  & $\mathbf{<.001}$ & 0.80 & $-0.02$ & 0.06 \\
    GazeController    & $-12.58$ & $\mathbf{<.001}$ & 0.47 & $-0.36$\dag
                      & 1.21     & 0.11             & 0.50 & $-0.21$ & $-0.16$ \\
    Hand              & -24.85    & $\mathbf{<.001}$ & 0.62 & 0.71\dag\dag
                      & -4.86    & 1                & 0.35 & -0.27 & -0.23 \\
    \bottomrule
  \end{tabular*}
  \label{tab:pairedztest-tostmanipulationwithnoconfirmationoperators}
  \vspace{-5pt}
\end{table*}

\begin{table*}[!t]
  \centering
    \caption{Paired Z-Test and TOST results for Placement, Search, and Scaling across modalities.}
  \small
  \setlength{\tabcolsep}{2.5pt}
  \renewcommand{\arraystretch}{0.9}
  \begin{tabular}{l ccc ccc ccc ccc ccc ccc}
    \toprule
    Modality 
      & \multicolumn{6}{c}{Placement} 
      & \multicolumn{6}{c}{Search}
      & \multicolumn{6}{c}{Scaling} \\
    \cmidrule(lr){2-7}\cmidrule(lr){8-13}\cmidrule(lr){14-19}
      & \multicolumn{3}{c}{Z-Test} & \multicolumn{3}{c}{TOST}
      & \multicolumn{3}{c}{Z-Test} & \multicolumn{3}{c}{TOST}
      & \multicolumn{3}{c}{Z-Test} & \multicolumn{3}{c}{TOST} \\
    \cmidrule(lr){2-4}\cmidrule(lr){5-7}
    \cmidrule(lr){8-10}\cmidrule(lr){11-13}
    \cmidrule(lr){14-16}\cmidrule(lr){17-19}
      & $Z$ & $p$ & $d$ & $t$ & $p$ & $SD$
      & $Z$ & $p$ & $d$ & $t$ & $p$ & $SD$
      & $Z$ & $p$ & $d$ & $t$ & $p$ & $SD$ \\
    \midrule
    Controller       
        & 53.26 & $\mathbf{<.001}$ & 1.52\dag\dag\dag 
        & 35.53 & 1 & 1.10
        & 4.78  & $\mathbf{<.001}$ & 0.14 
        & $-6.78$ & $\mathbf{<.001}$ & 0.59
        & 24.82 & $\mathbf{<.001}$ & 0.71\dag\dag 
        & 11.00 & 1 & 0.81 \\ 
    ControllerBlink  
        & 33.06 & $\mathbf{<.001}$ & 0.94\dag\dag\dag 
        & 24.13 & 1 & 1.31
        & 28.81 & $\mathbf{<.001}$ & 0.82\dag\dag\dag 
        & 19.57 & 1 & 0.76
        & 15.28 & $\mathbf{<.001}$ & 0.43\dag 
        & 7.08  & 1 & 1.13 \\ 
    GazeAirtap       
        & 1.03  & 0.30 & 0.03 
        & $-3.48$ & $\mathbf{<.001}$ & 1.38
        & 2.55  & $\mathbf{0.01}$ & 0.07 
        & $-2.91$ & $\mathbf{<.001}$ & 1.29
        & $-0.77$ & 0.44 & $-0.02$ 
        & 4.78  & $\mathbf{<.001}$ & 1.22 \\ 
    GazeController   
        & $-6.21$ & $\mathbf{<.001}$ & $-0.18$ 
        & 3.79  & $\mathbf{<.001}$ & 0.75
        & $-13.47$& $\mathbf{<.001}$ & $-0.39$\dag 
        & $-3.31$ & 1 & 0.69
        & $-7.61$ & $\mathbf{<.001}$ & $-0.22$\dag 
        & 1.38  & 0.08 & 0.82 \\ 
    Hand             
        & 2.14 & 0.03 & 0.06 
        & -11.07 & $\mathbf{<.001}$ & 0.51
        & $-70.85$& $\mathbf{<.001}$ & $-2.02$\dag\dag\dag 
        & $-52.85$& 1 & 0.39
        & -13.27 & $\mathbf{<.001}$ & -0.38\dag 
        & $2.42$ & $\mathbf{<.001}$ & 2.11 \\
    \bottomrule
  \end{tabular}
  \label{ManipulationPairedZWithoutConfirmationOperators}
    \vspace{-20pt}
\end{table*}

\begin{figure*}[t]
  \centering
  \includegraphics[width=0.75\textwidth]{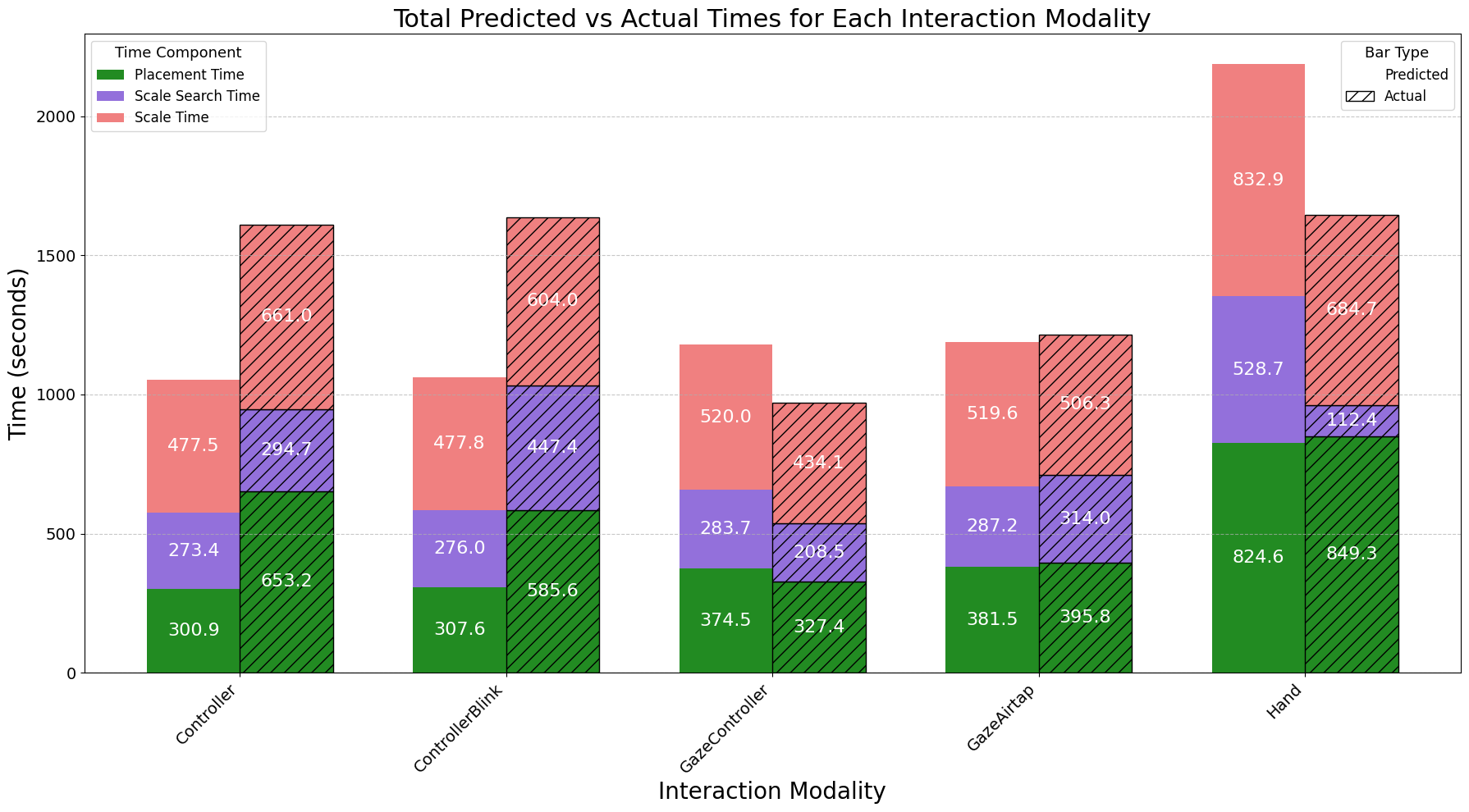}
  \vspace{-15pt}
  \caption{Total predicted vs. actual times without confirmation operators for each of the three phases of a manipulation trial.}
  \label{IndividualTotalPredictedVActual}
\vspace{-16pt}
\end{figure*}

One final set of paired Z tests was conducted on the data for each of the three phases of a trial (shown in Table~\ref{ManipulationPairedZWithoutConfirmationOperators} and Figure~\ref{IndividualTotalPredictedVActual}). This data looked at the $MT$ operator values only. 

Both Table \ref{ManipulationPairedZWithoutConfirmationOperators} and Figure \ref{IndividualTotalPredictedVActual} show that in the placement phase, \textsc{Controller} and \textsc{ControllerBlink} were not modeled well, with both modalities having large Z values and effect sizes. In the search phase, the Z values are smaller compared to the placement phase. However, \textsc{ControllerBlink} and \textsc{Hand} both have larger Z values with large effect sizes. Finally, in the scaling phase, the Z values are all smaller except for \textsc{Controller} which has a larger Z value along with a medium effect size. Through all three phases, \textsc{GazeController} and \textsc{GazeAirtap} have smaller Z values with smaller effect sizes. This pattern is also shown in Figure \ref{IndividualTotalPredictedVActual} where the actual and predicted bars for the modalities are close to one another. 

\section{Discussion}
\subsection{RQ1: Viability of Additive Models}
Our first research question asked: \textbf{``Is it feasible to use additive predictive models to predict the completion time of complex AR interactions using tasks similar to those found in real-world 3D user interfaces?"}

A 20\% difference between a user's actual and predicted interaction time has been the standard used to determine if a predictive model was ``good enough" in past GOMS literature \cite{cardKeystrokelevelModelUser1980, johnCumulatingScienceHCI1989, olsonGrowthCognitiveModeling1995}. Applying that standard suggests that our additive models worked well.  

Figure \ref{Totalvpredicted} shows that five of the six additive models performed well in the menu selection study: \textsc{Controller}, \textsc{GazeController}, \textsc{GazeAirtap}, \textsc{GazeDwell}, and \textsc{Hand}.  Although four of these modalities had delta times that were significantly different and the \textsc{Hand} modality was over the 20\% difference threshold, the effect sizes were small. The small effect sizes mean that there is likely little practical difference between the actual and predicted times, which the results from the TOST confirm, as shown in Table \ref{tab:pairedztest-tostmenuwithconfirmationoperators}. 

With the manipulation study data, we found that the \textsc{Controller}, \textsc{GazeController}, \textsc{GazeAirtap}, and \textsc{Hand} additive models produced total actual and predicted time values with a percentage difference under 20\% as shown in Figure \ref{TotalvpredictedManipulation}. Both \textsc{GazeController}, \textsc{GazeAirtap}, and \textsc{Hand} had negligible or small effect sizes, while the \textsc{Controller} modality was at the threshold between small and medium effect sizes (D = 0.50). The results from the TOST show that there is significant equivalence between the actual and predicted times for the \textsc{Controller}, \textsc{GazeController}, \textsc{GazeAirtap}, and \textsc{Hand} modalities. 

The \textsc{ControllerBlink} additive model had a percentage difference that was over 20\% in both the menu selection and manipulation studies. Considering the percentage differences, Z-scores, effect sizes, and TOST results for the \textsc{ControllerBlink} modality, it is more inaccurate compared to the other additive models. 

Ignoring the modalities with a percentage difference over 20\%, we find that the relative performance of the remaining additive models is quite good. As tables \ref{MenuAverageValues} and \ref{ManipulationAverageValues} show, the predicted and actual rankings nearly match exactly, with each table having two misplaced modalities. This is promising, showing that additive models of this sort can be useful in predicting which modalities/interfaces will outperform others. We found that with all modalities included, rankings often differed by at least one rank and sometimes differed by as much as two ranks. This means that the predictive capability of models should be taken into consideration when used for relative comparisons between input modalities as including poorly performing models can harm the rankings of the relative comparison.

One possible cause of the \textsc{ControllerBlink} additive model being inaccurate is participant fatigue from intentionally blinking their eyes to make selections. Similar effects have been reported in prior work \cite{luactivation21}, and it was fairly common for participants to express their displeasure regarding the \textsc{ControllerBlink} modality at some point during the study. This sentiment was also expressed in our NASA-TLX data with the \textsc{ControllerBlink} modality having the second largest physical demand and frustration values in the menu selection study (5.92 and 4.13 on a 10-point scale, respectively) and the largest physical demand value (5.55) and second largest frustration value (4.35) in the manipulation study\cite{hartDevelopmentNASATLXTask1988}.

Another way we approached testing the additive models for the manipulation task was by analyzing the data from the three phases of the trial individually to see if any one phase was more inaccurate. Looking at the data in Table \ref{ManipulationPairedZWithoutConfirmationOperators}, all of the controller modalities struggled to provide accurate predictions for the placement phase as indicated by their large Z-scores, large effect sizes, and their results from the TOST tests. During the search phase, only the \textsc{Controller} and \textsc{GazeAirtap} modalities performed well with smaller Z-scores, small or negligible effect sizes, and passing the TOST. During the scaling phase, all of the modalities did well with the only notable difference being that \textsc{Controller} had a medium effect size and that the p-value from the TOST results for the \textsc{GazeController} modality was just barely above the 0.05 significance level. One possible reason for the prediction inaccuracy in the placement phase is that it was the most complex phase. The complexity stems from the fact that the movement that occurs during that phase can be in one of two dimensions (along the horizontal axis or in and out of the depth axis) while the search phase only requires a constant amount of vertical movement, and the scaling phase only requires a variable amount of vertical movement. Participants may have developed proficiency in the latter two phases because they knew those phases would always require an upward movement.

Overall, our data shows that additive models are a feasible tool for predicting 3D interaction performance, provided that the underlying operators that comprise the additive model themselves provide accurate predictions for completion times, which we found to be the case for five of the six modalities tested across both studies. We found that the predictive capability of the five additive models was sufficient and provided accurate time predictions. We speculate that the \textsc{ControllerBlink} model may have also provided reasonable predictions, but may have suffered from issues with fatigue. Future work is needed to validate and improve the accuracy of KLM models on these techniques.

\subsection{RQ2: Accuracy of Past Predictive Models}
Our second research question asked: \textbf{``To what extent are existing models for MT validated when used to predict performance for individual components of more complex AR interactions?"}  

Looking at the data from the menu selection and manipulation studies, we see that the gaze-based models performed well. Figures \ref{Totalvpredicted} and \ref{TotalvpredictedManipulation} show that the percentage differences between the total actual and predicted time for \textsc{GazeController}, \textsc{GazeAirtap}, and \textsc{GazeDwell} are under 20\%, which would be considered an acceptable level of error for predictions \cite{cardKeystrokelevelModelUser1980, johnCumulatingScienceHCI1989, olsonGrowthCognitiveModeling1995}. All three modalities had small Z-scores, small effect sizes, and were shown to produce statistically equivalent predictions as shown in Tables \ref{tab:pairedztest-tostmenuwithnoconfirmationoperators} and \ref{tab:pairedztest-tostmanipulationwithnoconfirmationoperators}. 

The \textsc{Hand} $MT$ model performed well in the menu selection study. Figure \ref{Totalvpredicted} shows that the percentage difference is just over the 20\% threshold, however, the Z-scores, effect sizes, and results from the TOST show that the performance of the $MT$ model is acceptable. Figure \ref{TotalvpredictedManipulation} shows that the percentage difference for the \textsc{Hand} modality in the manipulation study is very low, with the total predicted and actual times being nearly equal. In the manipulation study, without considering the confirmation operators, the \textsc{Hand} $MT$ model fails the TOST equivalency test and has an effect size that borders on being large. Looking at the individual phases in Table \ref{ManipulationPairedZWithoutConfirmationOperators}, we see that the \textsc{Hand} $MT$ model has acceptable performance in both the placement and scaling phases; however, the performance in the search phase was poor and likely affected the overall results of the testing shown in Table \ref{tab:pairedztest-tostmanipulationwithnoconfirmationoperators}. The movement during the search phase (i.e., moving from the placement handle to the scaling handle) requires the participant to move a small distance upwards towards a large target, with no change in depth. This movement produces a very small $ID$ value. It is possible that the hand model, created by Machuca and Stuerzlinger\cite{barreramachucaEffectStereoDisplay2019}, was not created using trials with similarly small $ID$ values. As a result, the model produces poor predictions when $ID$ is very small.

We found that the predicted times for the confirmation operators were close to our observed times, with the largest difference between the actual and predicted confirmation operator value being the button-press confirmation operator in the menu selection study. This could have been a result of using different hardware, as Cabric et al.'s study used the HoloLens v1 with a clicker-style remote, while our work used a controller with a trigger button \cite{cabricPredictivePerformanceModel2021}. This also could have affected the confirmation operator time with the \textsc{GazeController} modality since its delta values varied from the predicted time as well ($\Delta$ = 120 ms and 121 ms). One interesting note is that the actual times for the blink operator were very similar to the predicted time from Caffier et al.'s work across both studies with the operator having a $\Delta$ value of 3 ms in the menu selection study and a $\Delta$ value of 68 ms in the manipulation study \cite{caffierExperimentalEvaluationEyeblink2003}. Even with tracking issues, the \textsc{GazeAirtap} confirmation operators were close to the predicted values with $\Delta$ values of 10 ms and 113 ms in the menu selection and manipulation studies, respectively. Future work should continue exploring the accuracy of the values by Cabric et al. \cite{cabricPredictivePerformanceModel2021} and gather device-agnostic values across a variety of HWDs.

\section{Limitations and Future Work}
Our work was not without limitations. First, our participant pool was insufficiently balanced in terms of gender. Future work should be conducted to see if a balanced participant pool affects the accuracy of additive models. Another limitation to our work is that the majority of our participants had prior experience with immersive technology. Future work should be conducted with a less experienced participant pool to see how the performance of additive models changes. 

\section{Conclusion}
In this work, we evaluated the validity of using additive models to predict the overall completion time of simple and complex interactions for 3D interfaces. We tested additive models by conducting two studies: one that had participants perform basic menu selection tasks and another where they manipulated a virtual cylinder across a tabletop to complete a multi-step manipulation task. We found that additive models work well in modeling both simple and complex interactions as long as the underlying components of an additive model themselves have accurate time predictions. Future work should explore using additive models to predict task completion time across a variety of 3D interfaces. 

\acknowledgments{
This research was funded by a grant from the U.S. Office of Naval Research. We would also like to thank our study participants for their contribution to our work.}

\bibliographystyle{abbrv-doi}

\bibliography{references}
\end{document}